# Non Equilibrium Green's Function Analysis of Double Gate SiGe and GaAs Tunnel FETs

Rahul Mishra[1], Bahniman Ghosh[1]

[1]Department of Electrical Engineering , Indian Institute of Technology, Kanpur. Kanpur, Uttar Pradesh, India. (208016)
email: rakaiit@gmail.com

**Abstract** – In recent past extensive device simulation work has already been done on TFETs. Various ways have been suggested to model TFETs. In our paper we look at one such particular way to model these devices. The Non equilibrium green's formalism has proved effective in modeling nano scale devices. We model complete SiGe and GaAs tunnel FET for the first time using the NEGF formalism, also taking acoustic phonon scattering into account. We analyze them on the grounds of I-V curve, $I_{on}$-$I_{off}$ ration and subthreshold slope. The poisson equation and the equilibrium statistical mechanical equation has been solved by providing the potential profile.

## I. Introduction

In recent years, studies on Tunnel FETs have proved them to be better than conventional MOSFETs as the former have steeper sub threshold swing, higher $I_{on}/I_{off}$ ratio, lower power consumption and their scaling is not limited by the quantum mechanical effect unlike the MOSFETs [1]-[10]. Tunnel FETs are basically low power devices. They work on the principle of band to band tunneling, from the valence band in source side to the conduction band in channel side [1]. The tunneling is initiated with the application of gate voltage, which lowers the tunneling distance between the two bands. In the past, studies have been done on tunnel FETs using device simulation softwares based on drift diffusion equations such as, medici, silvaco etc. As device sizes are scaled down, quantum effects become important and drift diffusion based simulators can no longer be relied upon for accurate results. A way to model quantum transport in such nano scale devices is the non equilibrium green's function (NEGF) formalism suggested in [11]. In this formalism, Poisson equation and equilibrium statistical mechanics equations are solved self consistently. In this work, we model complete SiGe and GaAs TFETs using the NEGF formalism. We extract the potential profile using device simulation software for solving the NEGF equations. These TFETs are then analyzed on the grounds of their current voltage characteristics, their Ion/Ioff ratios and their sub threshold slopes. GaAs TFET is found to be better in terms of the $I_{on}/I_{off}$ ratio. However SiGe TFETs are found to have better subthreshold swing. To the best of our knowledge, no such studies have been reported so far in the literature

## II. Simulation Methodology

We model a TEFT using the NEGF formalism which we describe briefly below. For a complete description of the NEGF technique we refer the reader to [11]. For a device in equilibrium we first identify a suitable Hamiltonian, that provides adequate description of the isolated device. On connecting the device to source and drain contacts and applying appropriate voltages, charge is transferred in and out of the device The potential, *U(r)*, is calculated self-consistently with the charge profile. The solver iterates between the Poisson equation that gives us the potential (U(r)), for a given electron density *n(r)* relative to that required for local charge neutrality

$$\nabla . (\varepsilon \nabla U) = q^2 [N_D - n] \quad (1)$$

And the electron density n(r) for a given potential profile U(r) is obtained using the law of equilibrium statistical mechanic is given by

$$n(r) = \sum_\alpha |\psi_\alpha(r)|^2 f_0(\varepsilon_\alpha - \mu) \quad (2)$$

where Ψ denotes the wavefunction obtained from the Schrödinger equation

$$[H+U]\psi_\alpha(r) = \varepsilon_\alpha \psi_\alpha(r) \quad (3)$$

According to the Fermi function

$$f_0(E-\mu) = (1+\exp[(E-\mu)/k_B T])^{-1} \quad (4)$$

where µ is the Fermi level. We will in brief discuss various equations used for obtaining the current through the device. Details are mentioned in [11]. We use the model for dissipative quantum transport described below

$$G^n = G \sum^{in} G^+ \quad (5)$$

where $G^n$ is the correlation function and

$$G = [EI - H_0 - U - \Sigma] \quad (6)$$

is the Green's function, $E$ is the energy, $H_0$ is the Hamiltonian, $\Sigma$ is the total self energy including the effects of scattering and the contacts and $U$ is the electrostatic potential.

The current at any terminal $i$ can be calculated using

$$I_i = (q/\hbar) \int_{-\infty}^{\infty} dE \, \tilde{I}_i(E)/2\pi \quad (7)$$

where $q$ is the charge of an electron, $\hbar$ is the Planck's constant divided by $2\pi$, and

$$\tilde{I}_i = \text{Trace}[\Sigma_i^{in} A] - \text{Trace}[\Gamma_i G^n] \quad (8)$$

where

$$A = i[G - G^+] \quad (9)$$

and

$$\Gamma_i = i[\Sigma_i - \Sigma_i^+] \quad (10)$$

$\Sigma_i^{in}$ is the inscattering function.

In the literature, ballistic simulations have been performed extensively using the NEGF formalism. In this work, we also take into account phonon scattering. To keep the analysis simple we limit ourselves to acoustic phonon scattering which we model by a deformation potential. The self energy, $\Sigma_s$, for the acoustic phonon scattering is given by, [12]

$$\Sigma_s(E) = K_a G(E) \quad (11)$$

Here $K_a$ is the coupling constant given by, [12]

$$K_a = \frac{D_a^2 K_B T}{\rho v_a^2 a^3} \quad (12)$$

where $D_a$ is the deformation potential, $K_B$ is the Boltzmann constant, T is the temperature, $\rho$ is the crystal density, $v_a$ is the acoustic phonon velocity and $a$ is the lattice constant. $\Sigma_s(E)$ and $G(E)$ are calculated iteratively till they converge.

## III. Device parameters

We model a double gate tunnel FET using NEGF. The substrate is P doped with carrier concentration of $10^{16}$ cm$^{-3}$. The source region is P doped with density of $10^{20}$ cm$^{-3}$. The drain region is N$^+$ doped with density of $10^{18}$ cm$^{-3}$. Metallic gate with workfunction 4.3 ev is used in case of N channel tunnel FET. Table I shows rest of the parameters. Acoustic deformation potential values are given in table II.

**Table I.** Device Parameter of NTFET used in simulation

| | |
|---|---|
| Source Doping (atoms/cm$^3$) | $10^{20}$ |
| Drain Doping (atoms/cm$^3$) | $10^{18}$ |
| Substrate Doping (atoms/cm$^3$) | $10^{16}$ |
| Channel Length | 20nm |
| Dielectric thickness | 3 nm |
| Body thickness | 10 nm |
| Gate work function | 4.3 eV |

**Table II.** Acoustic Deformation potential

| | |
|---|---|
| Si | 9 eV |
| Ge | 11 eV |
| GaAs | 11 eV |

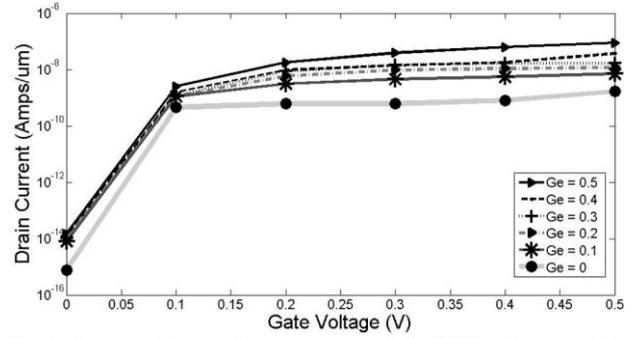

Fig. 1. Current – Voltage Characteristic for SiGe TFETs. $V_{DS}$ = 0.05 V.

## IV. Results

*A. Current-voltage characteristics*

Fig. 1 compares the $I_d$-$V_g$ curves of complete SiGe tunnel FETs for various Ge mole fractions (0 to 0.5). Clearly increasing Ge mole fraction decreases the tunneling width due to decreasing band gap and hence increases the drain current. On current (at $V_{gs}$= 0.5 V, $V_{ds}$= 0.05 V) is $8.8816 \times 10^{-8}$ A/µm for Ge mole fraction of 0.5 compared to $1.6883 \times 10^{-9}$ A/µm for pure silicon. For SiGe we use conductivity effective mass of $0.26 m_0$ [13]. Fig.2 shows the I-V curve for GaAs tunnel FET. The conductivity effective mass used in case of GaAs is $0.067 m_0$ [14]. On current (at $V_{gs}$= 0.5 V, $V_{ds}$= 0.05 V) is $3.1778 \times 10^{-7}$ A/µm and off current is $1.6359 \times 10^{-14}$ A/µm for the GaAs TFET. Comparing the two we see that I-V curve is steep in case of SiGe as compared to GaAs TFETs.

*B. $I_{on}$-$I_{off}$ ratio*

We take a look at the $I_{on}$ – $I_{off}$ ratio of SiGe and GaAs TFETs. Among SiGe's the best $I_{on}$-$I_{off}$ ratio is for the SiGe TFET with $I_{on}$ - $I_{off}$ ratio of $5.9 \times 10^6$. $I_{on}$–$I_{off}$ ratio for the GaAs TFET comes out to be $1.94 \times 10^7$. Thus GaAs TFET perform better than SiGe TFETs when it comes to comparing the $I_{on}$-$I_{off}$ ratio.

*C. Subthreshold Slope*

The subthreshold swing is calculated using the formulae

$$S = \frac{V_t - V_{off}}{\log I_{vt} - \log I_{off}}$$

Fig. 3 shows the subthreshold slopes of SiGe and GaAs TFETs. As clear from the graph SiGe TFETs have better subthreshold slope than the GaAs TFET. The reason for.

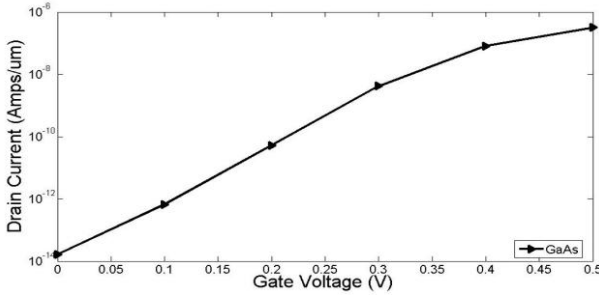

Fig. 2. Current – Voltage Characteristic for GaAs TFET. $V_{DS}$ = 0.05 V.

this becomes clear when we look at the I-V curve of the two devices, SiGe has steeper curve compared to GaAs and hence better subthreshold swing. High $I_{on}/I_{off}$ ratio of GaAs does not necessarily mean it will have better subthreshold swing, as subthreshold swing is defined as voltage required for a decade rise in current near the threshold region, which is independent of the $I_{on}/I_{off}$ ratio. SiGe TFET with Ge mole fraction of 0.5 has a subthreshold slope of 18.64 mV/Dec, GaAs TFET has subthreshold slope of 57.1 mV/Dec. Thus SiGe TFETs out perform the GaAs TFETs when it comes to the subthreshold slope.

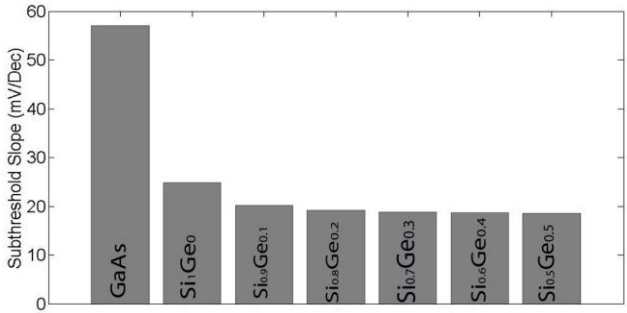

Fig. 3. Subthreshold Slope for different TFETs. $V_{DS}$ = 0.05 V

## IV. Conclusion

In this work, we perform quantum transport simulation studies of SiGe and GaAs TFETs using the NEGF formalism. We observe that the NEGF formalism gives good results for the TFETs. GaAs TFET performs better when we consider the $I_{on}$-$I_{off}$ ratio, however SiGe TFETs score better when it comes to subthreshold slope. We note that NEGF is a good method to model nano scale devices, and more work in this area will be helpful to obtain insight into the nature of quantum transport in these devices